\documentclass[twocolumn,letter]{jpsj2} 
\usepackage{amsmath}
\usepackage{amsfonts}
\usepackage{amssymb}
\usepackage{graphicx}
\usepackage{dcolumn}
\usepackage{bm}

\title{Coherent Behavior and Nonmagnetic Impurity Effects\\of the Spin Disordered State in NiGa$_2$S$_4$}

\author{Yusuke \textsc{Nambu}$^{1}$, 
Satoru \textsc{Nakatsuji}$^{1}$ and 
Yoshiteru \textsc{Maeno}$^{1,2}$}

\inst{$^{1}$Department of Physics, Kyoto University, Kyoto 606-8502\\
$^{2}$International Innovation Center, Kyoto University, Kyoto 606-8501}

\abst{Nonmagnetic impurity effects of the spin disordered state in the triangular antiferromagnet 
NiGa$_2$S$_4$ was studied through magnetic and thermal measurements for Ni$_{1-x}$Zn$_x$Ga$_2$S$_4$ 
$\left(0.0\le x\le 0.3\right)$. Only 1 \% substitution is enough to strongly suppress the coherence 
observed in the spin disordered state. However, the suppression is not complete and the robust feature of 
the $T^2$ dependent specific heat and its scaling behavior with the Weiss temperature indicate the 
existence of a coherent Nambu-Goldstone mode. Absence of either conventional magnetic order or bulk spin 
freezing suggests a novel symmetry breaking of the ground state.}

\kword{triangular lattice, NiGa$_2$S$_4$, spin disordered state, Nambu-Goldstone mode}

\begin{document}
\maketitle

Geometrically frustrated magnets have attracted interest because of the possible emergence of novel 
magnetic phases at low temperatures by suppressing conventional magnetic order. Among such magnets, one 
of the simplest and most fundamental structures is the two-dimensional (2D) triangular lattice that has 
a single magnetic ion per unit cell. The 2D triangular lattice antiferromagnets (AFMs) have long 
attracted interest since a quantum spin disordered state was theoretically proposed more than three 
decades ago \cite{RVB1}. While it is believed that the triangular AFM with nearest-neighbor coupling 
exhibits the so-called 120$^{\circ}$ order \cite{Huse,Bernu,Capriotti}, recent theories suggest that the 
exchange interaction beyond nearest-neighbor such as longer range and multiple-spin exchange interactions 
may lead to a quantum disordered state \cite{Morita,ML}. For experiment, only a few quasi-2D triangular 
AFMs with low spin, i.e. $S \le 1$, have been reported, such as a solid $^{3}$He thin film \cite{He} and 
an organic material with a distorted triangular lattice \cite{organic}. 

Recently, a new quasi-2D $S=1$ triangular AFM NiGa$_2$S$_4$ has been discovered as the first example of 
a bulk low-spin AFM with an exact triangular lattice \cite{Nigas}. Interestingly, despite 
antiferromagnetic (AF) coupling of $\sim 80$ K, neither long-range order nor bulk spin freezing has 
been detected down to 0.35 K. Instead, magnetic, thermal and neutron diffraction measurements indicate 
the formation of a gapless spin disordered state below about 10 K. The quadratic temperature dependence 
of the specific heat, as well as the temperature independent behavior of the susceptibility at 
$T\rightarrow 0$ reveal coherence of a gapless linearly dispersive mode in two dimensions. Such a spin 
disordered state is presumably sensitive to perturbation such as impurity doping. In order to 
clarify the mechanism of the formation of the unusual low temperature state, it is highly important to 
examine experimentally the stability against impurities of the spin disordered state and its gapless 
linearly dispersive mode.

In this letter, we report the nonmagnetic impurity effects of NiGa$_2$S$_4$ based on the magnetic and 
thermal measurements for the Zn substituted insulating materials Ni$_{1-x}$Zn$_x$Ga$_2$S$_4$. Only 1 \% 
is enough to substantially suppress the coherence of the spin disordered state in NiGa$_2$S$_4$. However, 
the Zn substitution never completely suppresses the coherence, but induces a ``defect'' component of 
weakly coupled $S=1$ spin up to a few percent, which freezes at low temperatures. The robust feature of 
the quadratic temperature dependence of the magnetic specific heat, and its scaling with the Weiss 
temperature indicate the existence of the Nambu-Goldstone mode of a gapless 
linearly dispersive type. Absence of either conventional magnetic order or bulk spin freezing suggests 
a novel symmetry breaking of the ground state.

The polycrystalline samples of Ni$_{1-x}$Zn$_x$Ga$_2$S$_4$ are synthesized by annealing of Ni, Zn, Ga 
and S elements in evacuated silica tube at 850 $\sim$ 900 $^{\circ}$C. In order to obtain a homogeneous 
mixture of Ni and Zn, we first ground the same molar amount of Ni and Zn powder, and repeated the 
dilutions by adding equal molar amount of Ni powder each time until we obtain an appropriate Zn 
concentration. Powder x-ray diffraction analysis at room temperature on our polycrystalline samples 
indicates single phases with the same trigonal structure as NiGa$_2$S$_4$ up to $x = 0.3$. Our 
preliminary neutron diffraction measurements for $x = 0,\ 0.1,\ 0.25$ confirmed both the trigonal 
structure with P$\overline{3}$m1 symmetry and the substitution of Zn at the Ni site at the same 
concentration as the nominal one within an error of 1 \% \cite{Robin}. For $x\ge 0.4$, two phases 
coexistence of the trigonal and tetragonal phases is found because of a phase transition to the 
tetragonal phase of ZnGa$_2$S$_4$. In this letter, we focus on the physical properties of the trigonal 
phase region $(0 \le x \le 0.3)$. 

DC magnetization $M$ was measured between 1.8 and 350 K under 
magnetic fields up to 7 T using a SQUID magnetometer. The specific heat $C_P$ was measured by thermal 
relaxation method down to 0.35 K. In order to estimate the lattice contribution, 
$C_\mathrm{L}$, we measured $C_P$ of the isostructural nonmagnetic analogue ZnIn$_2$S$_4$ and get the 
thermal variation of Debye temperature $\theta_\mathrm{D}\left(T\right)$ using Debye equation 
\cite{Debye}. $\theta_\mathrm{D}\left(T\right)$ of Ni$_{1-x}$Zn$_x$Ga$_2$S$_4$ is then estimated by 
multiplying a scaling factor according to $\theta_\mathrm{D}\left(T\right)\propto M_0^{-1/2}V_0^{-1/3}$, 
where $M_0$ and $V_0$ are molar mass and volume, respectively. $C_\mathrm{L}$ is thus estimated by 
converting back this scaled $\theta_\mathrm{D}\left(T\right)$ into specific heat. 

\begin{figure}[tb]
\begin{center}
\includegraphics[scale=0.47]{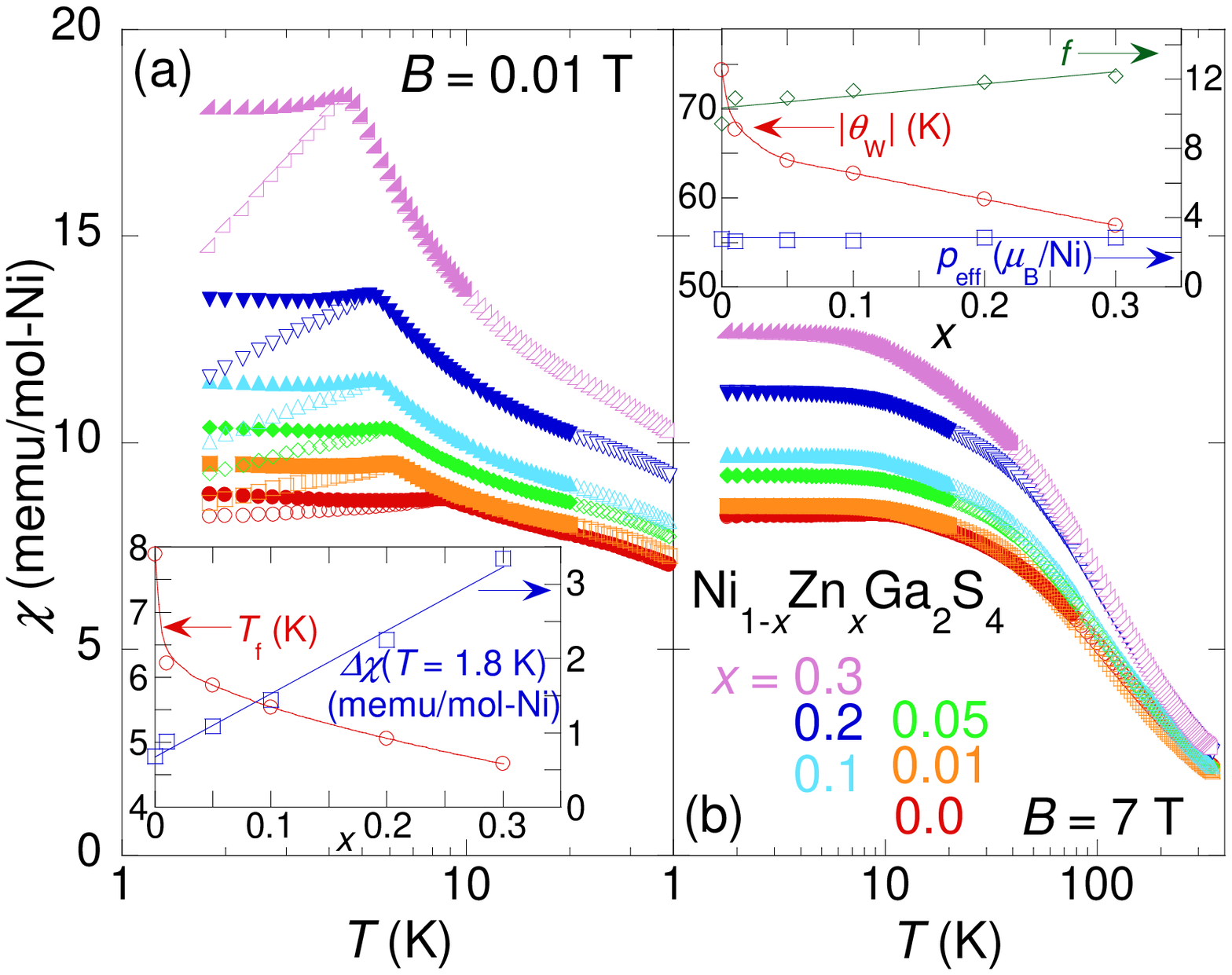}
\end{center}
\vspace{-0.0in}
\caption{Temperature dependence of the susceptibility $\chi\left(T\right)$ for both FC (solid) and 
ZFC (open) sequences under (a) $B=0.01 $ T and (b) 7 T. The insets show (a) the $x$ dependence of the 
freezing temperature $T_{\mathrm{f}}$ and $\varDelta\chi\left(T=1.8\ \mathrm{K}\right)$ under $B=0.01$ T, 
and (b) the effective moment $p_{\mathrm{eff}}$ per Ni ion, the absolute value of the Weiss temperature 
$|\theta_{\mathrm{W}}|$ and the frustration parameter $f$. The blue horizontal line gives 
$p_{\mathrm{eff}}=2.84$ expected for $S=1$.}
\label{magnetization}
\end{figure}

The temperature dependence of the susceptibility $\chi\left(T\right)\equiv M\left(T\right)/B$ under fields 
$B=0.01$ and 7 T is presented in Fig. 1. Under 0.01 T, samples of all concentrations show the bifurcation 
between field cooled (FC) and zero-field cooled (ZFC) data below freezing temperature $T_\mathrm{f}$ 
(Fig. 1(a)). For the pure NiGa$_2$S$_4$, the freezing effect is so tiny that it is not the bulk spins 
that freeze, but nearly 300 ppm defect spins due to the imperfection of the samples including surface 
\cite{Nigas}. The Zn substitution increases the difference between FC and ZFC data at 1.8 K, 
$\varDelta\chi\left(T=1.8\ \mathrm{K}\right)$, linearly with the Zn concentration $x$ 
(inset of Fig. 1(a)). Under 7 T, no hysteresis between FC and ZFC data was observed down to 1.8 K 
(Fig. 1(b)). 

Above 150 K, the susceptibility data for all concentration follow the Curie-Weiss law: 
$\chi\left(T\right)=C/\left(T-\theta_\mathrm{W}\right)$. The effective moment $p_{\mathrm{eff}}$ per 
Ni ion estimated from the Curie constant $C$ is close to the value expected for $S=1$ 
(2.83 $\mu_{\mathrm{B}}$). This confirms that Zn substitutes Ni by the same amount as the nominal 
concentration. The Weiss temperature $\theta_\mathrm{W}$ is found negative and AF. As in the insets of 
Fig. 1(a) and 1(b), the absolute values of both $T_\mathrm{f}$ and $\theta_\mathrm{W}$ show a drastic 
change at $x \sim 0$, indicating the strong sensitivity of the bulk disordered state of NiGa$_2$S$_4$ 
to the impurity. Further substitution systematically decreases both $T_{\mathrm{f}}$ and 
$|\theta_{\mathrm{W}}|$ in a similar fashion. 

\begin{figure}[tb]
\begin{center}
\includegraphics[scale=0.47]{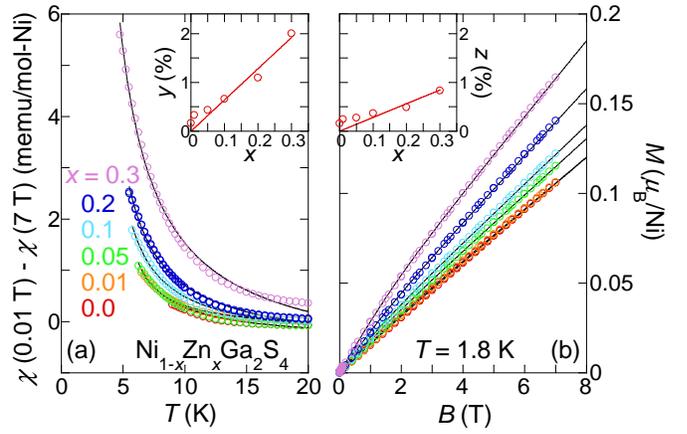}
\end{center}
\vspace{-0.0in}
\caption{(a) Difference between the susceptibility under $B=0.01$ T and 7 T fitted by 
$\chi_0+yC_{S=1}/\left(T-\theta\right)$. The inset shows its fitting result, $y$ vs. $x$. (b) 
Magnetization $M$ vs. the field $B$ fitted by eq. (\ref{MHtc}) in the text. The inset shows its fitting 
result, $z$ vs. $x$.}
\label{two-component}
\end{figure}

In order to estimate the amount of Ni spins affected by the Zn substitution, we performed a 
two-component analysis on the magnetization results. Here, the two components are the bulk spins most 
likely forming the spin disordered state similar to NiGa$_2$S$_4$, and ``defect spins'' induced by the 
Zn substitution. Comparing the results under 0.01 T (Fig. 1(a)) and 7 T (Fig. 1(b)), one notices that the 
low-$T$ susceptibility becomes significantly field dependent with $x$, while NiGa$_2$S$_4$ exhibits nearly 
field-independent susceptibility. This suggests that the defect spins are much more weakly coupled than 
the bulk spins with $|\theta_{\mathrm{W}}| = 80$ K, and can be fully polarized under 7 T. In order to 
confirm this, the difference between $B=0.01$ T and 7 T are fitted by 
$\chi_0+yC_{S=1}/\left(T-\theta\right)$ in the range of $T_{\mathrm{f}} \le T \le 20$ K, where $ C_{S=1}$ 
is the Curie constant of 1.0 (emu K/mol) for $S=1$ and a fitting parameter $y$ gives the defect spin 
fraction. All curves are fitted as shown in Fig. 2(a). The Weiss temperature for defect spins, $\theta$, 
was only about one half of $T_{\mathrm{f}}$ and much smaller than $|\theta_{\mathrm{W}}| $, confirming 
that the defect spins are weakly coupled. $\chi_0$ is negligibly small $\sim 10^{-4}$ emu/mol. 
The fraction of the defect spins, $y$, increases linearly with $x$. 

We also performed the two-component analysis on the field dependence of the magnetization 
$M\left(B\right)$. For the pure NiGa$_2$S$_4$, the field dependence data ($0\le B\le 7$ T) exhibit a 
linear increase at 1.8 K. However, non-linear Brillouin function type component appears and increases 
with the Zn substitution (Fig.2(b)) because of the weakly coupled defect spins. In order to estimate the 
amount of the defect spins, we performed the fitting using the following equation in the range of 
$0 \le B \le 7$ T,
\begin{align}
M\left(B,x\right)=\frac{dM}{dB}\bigg|_{B=7\mathrm{T}}B
+zg\mu_{\mathrm{B}}SB_S\left(\frac{g\mu_\mathrm{B}SB}{k_\mathrm{B}T}\right). 
\label{MHtc}
\end{align}
Here, we assume that the susceptibility for bulk spins is insensitive to field as in the pure 
NiGa$_2$S$_4$ and is given by $\frac{dM}{dB}$ at 7 T where the defect spins are polarized. $B_S$ in the 
second term represents the Brillouin function for $S = 1$ defect spins. 
The eq. (\ref{MHtc}) fits the experimental results well as shown in Fig 2 (b). 
The fraction of the defect spins, $z$, increases 
linearly with $x$. The above two types of the analyses consistently 
give the same order of the defect spin concentration that are proportional to $x$. This indicates that 
the nonmagnetic impurity substitution induces $S=1$ weakly coupled defect spins that most likely comes 
from Ni spins adjacent to Zn ions. Only a few percent of Ni site appears to induce the defect spins. 
Given the low temperature hysteresis $\varDelta\chi\left(T=1.8\ \mathrm{K}\right)\propto x$, these results 
indicate that {\it it is not the bulk spins but the defect spins that freeze}.

\begin{figure}[tb]
\begin{center}
\includegraphics[scale=0.5]{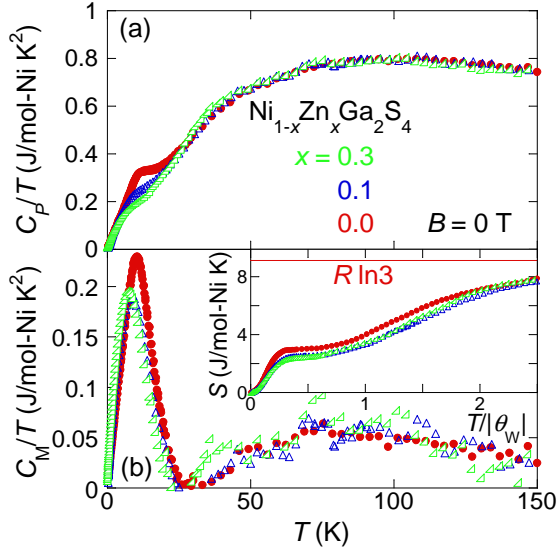}
\end{center}
\vspace{-0.0in}
\caption{$T$ dependence of (a) the total specific heat divided by temperature, $C_P/T$, (b) the 
magnetic specific heat devided by temperature, $C_{\mathrm{M}}/T$. The inset shows the magnetic entropy 
$S$ as a function of $T/|\theta_{\mathrm{W}}|$. The horizontal line indicates $S = R\ln3$.}
\label{specific heat}
\end{figure}

The absence of bulk spin freezing as well as spin order was confirmed by specific heat measurements 
under $B=0$ T (Fig. 3(a)). The $T$ dependence of the specific heat divided by $T$, 
$C_P/T \left(T\right)$, shows only broad anomalies, indicating that conventional AF order is still 
suppressed in Ni$_{1-x}$Zn$_x$Ga$_2$S$_4$ and that the interacting spin remain disordered in the low-$T$ 
limit. 
In order to probe the magnetic density of excited states, the $T$ dependence of the magnetic part 
of the specific heat $C_\mathrm{M}\left(T\right)$ was estimated by subtracting the lattice part 
$C_{\mathrm{L}}(T)$ from $C_P\left(T\right)$ (Fig. 3(b)). All $x \ne 0$ compounds have a double-peak 
structure, similar to NiGa$_2$S$_4$ ($x=0$). One is the broader peak centered around 
$T \sim |\theta_{\mathrm{W}}|$, and the other is the prominent rounded peak at the temperature close to 
$T_{\mathrm{f}}$.

Significantly, $C_{\mathrm{M}}/T$ for all $x$ shows a $T$ linear dependence at $T \rightarrow 0$.
This is in sharp contrast with $T$ independent $C_{\mathrm{M}}/T$ due to the local nature of spin 
fluctuations in canonical spin glasses. This indicates absence of spin glass state in the bulk, but the 
coherent propagation of a gapless and linearly dispersive mode in two dimensions.

With a 2D gapless linearly dispersive mode with a coherent propagation limited up to a length scale $L_0$ 
at $T=0$, the specific heat deviates from the low temperature asymptotic form as
\begin{align}
\frac{C_{\mathrm{M}}\left(T\right)}{R}=\frac{C_0}{R} 
+\frac{3\sqrt{3}\zeta\left(3\right)}{2\pi}\left(\frac{ak_{\mathrm{B}}T}{\hbar D}\right)^2, \label{cl}
\end{align}
for $hD/L_0 k_{\mathrm{B}}\ll T\ll |\theta_{\mathrm{W}}|$, where 
$C_0=-\left(\sqrt{3}\pi/2\right)\left(a/L_0\right)^2R$ and $\zeta\left(3\right)=1.202$ \cite{Nigas,SCGO}. 
Here $a$ is the lattice constant, and $D$ the spin stiffness constant. 

For NiGa$_2$S$_4$, $C_{\mathrm{M}}$ shows quadratic $T$ dependence with $C_0 = 0.0(2)$ mJ/mole K and this 
indicates infinitely long $L_0$ with the lower bound $\simeq 130$ nm, which is far beyond the two-spin 
correlation length $\xi \sim 2.5$ nm determined by the neutron scattering \cite{Nigas}. With the 
substitution of Zn, however, $L_0$ sharply decreases at $x \sim 0$ (inset of Fig. 4). This shows the 
coherence of the pure NiGa$_2$S$_4$ is fragile and can be easily suppressed by a few percent impurity. 
On the other hand, the spin stiffness constant shows the continuous decrease with $x$ (inset of Fig. 4). 
For ordinary AFMs that order at $T \sim |\theta_{\mathrm{W}}|$, the stiffness constant $D_0$ can be 
estimated by the relation: 
$D_0^2 \approx \left(3\sqrt{3}\zeta\left(3\right)/4\pi\right)\left(ak_{\mathrm{B}}\theta_{\mathrm{W}}/\hbar\right)^2/\ln\left(2S+1\right)$ \cite{Nigas,SCGO}. 
The observed $D$ is between 0.65 and 0.85 km/sec, nearly three times smaller than the expected 
$D_0 = 2.3$ - 3.0 km/sec. This softening comes from the magnetic frustration, and indicates the spectral 
weight downshift. Further evidence for the downshift is found in the magnetic entropy $S$ 
(inset of Fig. 3(b)), obtained by integration of $C_{\mathrm{M}}/T$. The observed intermediate 
temperature plateau of $S$ prior to high-$T$ saturation at $\sim R\ln3$ amounts to $1/4 \sim 1/3$ of 
$R\ln3$, indicating the highly degenerate low-$T$ states. 

\begin{figure}[tb]
\begin{center}
\includegraphics[scale=0.48]{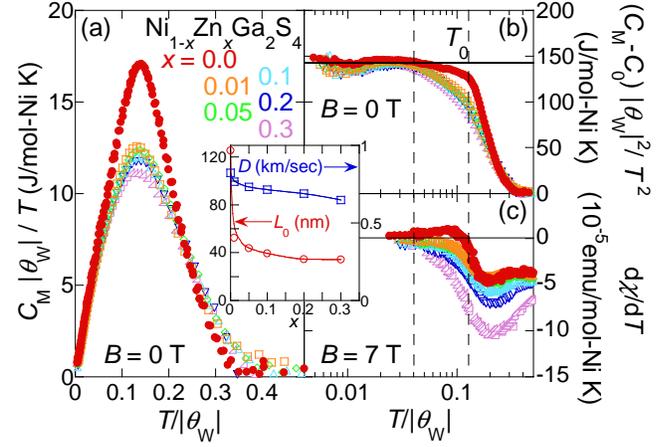}
\end{center}
\vspace{-0.0in}
\caption{$T/|\theta_{\mathrm{W}}|$ dependence of (a) $C_{\mathrm{M}}|\theta_{\mathrm{W}}|/T$, (b) 
$(C_{\mathrm{M}}-C_0)|\theta_{\mathrm{W}}|^2/T^2$, and (c) the derivative of the susceptibility 
$d\chi/dT$ under $B=7$ T. $C_0$ is defined for $hD/L_0 k_{\mathrm{B}}\ll T\ll |\theta_{\mathrm{W}}|$. 
The vertical dashed lines indicate 4 \% of $|\theta_{\mathrm{W}}|$ and $T_0 \simeq$ 13 \% of 
$|\theta_{\mathrm{W}}|$. The inset shows the $x$ dependence of the coherent scale $L_0$ and the 
stiffness constant $D$. $C_0 = 0.0(2)$ mJ/mole K for NiGa$_2$S$_4$ indicates infinitely long $L_0$ and 
its lower bound is plotted.}
\label{Weiss temp. normalize}
\end{figure}

Interestingly, the observed systematic change of the specific heat and susceptibility as a function of 
$x$ roughly scales with the Weiss temperature. When the system is dominated by the single scale 
$|\theta_{\mathrm{W}}|$, the molar entropy should take the form 
$S = f\left(T/|\theta_{\mathrm{W}}|\right)$. Indeed, $S$ presented in the inset of Fig. 3(b) shows 
such a scaling as a function of $T/|\theta_{\mathrm{W}}|$, except one for NiGa$_2$S$_4$. Then, the 
specific heat should be given by $T/|\theta_{\mathrm{W}}|$ times a scaling function of 
$T/|\theta_{\mathrm{W}}|$. In order to check this, we plot $C_{\mathrm{M}}|\theta_{\mathrm{W}}|/T$ vs. 
$T/|\theta_{\mathrm{W}}|$ in Fig. 4(a). Surprisingly, all the curves except the one for the pure 
NiGa$_2$S$_4$ overlaps on top of each other. Moreover, including NiGa$_2$S$_4$, the peak temperature of 
$C_{\mathrm{M}}|\theta_{\mathrm{W}}|/T$ and the initial slope as a function of $T/|\theta_{\mathrm{W}}|$ 
are constant, indicating that the peak temperature is proportional to $|\theta_{\mathrm{W}}|$, and $T^2$ 
constant of $C_{\mathrm{M}}$ is proportional to $|\theta_{\mathrm{W}}|^{-2}$. In order to further check 
this, $(C_{\mathrm{M}}-C_0)|\theta_{\mathrm{W}}|^2/T^2$ is plotted against $\ln (T/|\theta_{\mathrm{W}}|$) 
in Fig. 4(b). All the results including the one for NiGa$_2$S$_4$ (with $C_0=0.0\left(2\right)$ mJ/mole K) 
starts to have the same constant value below the temperature of $T/|\theta_{\mathrm{W}}| \sim 0.04$. 
While the lower $T$ limit to define $C_0$, $hD/L_0 k_{\mathrm{B}}$ corresponds to 
$T/|\theta_{\mathrm{W}}| = 0.015(4)$ for the Zn substituted samples, the eq. (\ref{cl}) roughly describes 
$C_{\mathrm{M}}$ down to the lowest $T$ given by $T/|\theta_{\mathrm{W}}| \sim 0.004$, much lower than 
the limit.

On the other hand, the scaling behavior in the susceptibility is not so clear as in the specific heat 
results probably because the susceptibility is more sensitive to defect spins at low temperatures. 
However, we note the three important characteristic temperatures that scale with $|\theta_{\mathrm{W}}|$. 
(1) Below 4 \% of $|\theta_{\mathrm{W}}|$, the susceptibility becomes constant, as 
indicated by the vanishing $T$ derivative, $d\chi/dT$ (Fig 4(c)).
(2) The minimum of $d\chi/dT$ appears at nearly the same scaled temperature of 
$T/|\theta_{\mathrm{W}}| \sim 0.2$ (Fig 4(c)).
(3) $T_{\mathrm{f}}$ scales with $|\theta_{\mathrm{W}}|$, so the frustration parameter 
$f\equiv|\theta_{\mathrm{W}}|/T_{\mathrm{f}}$ keeps almost a constant value around 10, indicating the 
significant magnetic frustration even with the Zn substitution (inset of Fig. 1(b)). 

Despite the above scaling behavior with the Zn substitution, 
only the pure NiGa$_2$S$_4$ exhibits the 
following two noticeable anomalies at the crossover temperature to the low-$T$ asymptotic behavior. 
(1) Not following a scaling function for the Zn substituted systems, only NiGa$_2$S$_4$ exhibits a 
broad kink in $(C_{\mathrm{M}}-C_0)|\theta_{\mathrm{W}}|^2/T^2$ at $T_0 \simeq$ 13 \% of 
$|\theta_{\mathrm{W}}| \simeq 10$ K, below which nearly $T^2$ dependent $C_{\mathrm{M}}$ appears.
(2) Only for NiGa$_2$S$_4$, $d\chi/dT$ crosses zero at the same characteristic temperature $T_0$, 
below which the nearly constant $\chi$ appears. Interestingly, this low temperature crossover behavior 
in the pure NiGa$_2$S$_4$ can be easily suppressed only by 1 \% Zn, similar to the rapid loss of the 
coherence scale $L_0$.

The Zn substitution substantially but never completely suppresses $L_0$ and $T^2$ dependent specific 
heat. The observed robust scaling of the low temperature behavior, especially the $T^2$ form of the 
specific heat, strongly suggests the existence of the Nambu-Goldstone mode, which has a gapless and 
linearly dispersive character and a $T^2$ coefficient scaling with $|\theta_{\mathrm{W}}|^{-2}$ 
\cite{SCGO}. Generally, a Nambu-Goldstone mode appears with a broken symmetry in comparison with its 
high-$T$ phase. Therefore, in our case, the absence of the evidence for conventional long-range order 
in thermodynamics and neutron results points to a novel magnetic order in two dimensions. 

One candidate for such a unconventional long-range order without long-range two-spin correlation is a spin 
nematic phase \cite{nematic}. This phase may be classified as a spin liquid because the static site average of 
spin is zero, while its magnetic quadrupole correlation is long-ranged. If the transition occurs at 
$T = T_0$, spin fluctuations should show a critical slowing down, which may well lead to the freezing of 
impurity spins at $T_{\mathrm{f}}$ close to $T_0$. Recently, the mean-field type calculations by 
Tsunetsugu and Arikawa have shown that a spin nematic order can be indeed stable on a 2D triangular 
lattice, and generates a Nambu-Goldstone mode of the gapless linearly dispersive type \cite{Tsunetsugu}. 
In our case, the strong two-dimensionality of the spin interactions may suppress the three-dimensional 
character of the order and instead promote a rapid development of the correlation at a crossover 
temperature $T_0$, resulting in the broad feature of the specific heat near $T_{\mathrm{f}}$. Furthermore, 
the orthogonal spin nematic order with three sublattices has no geometrical 
frustration on the triangular lattice \cite{Tsunetsugu}, and may be robust against the substitution of a 
sizable amount of nonmagnetic impurities.

Another candidate is a Kosterlitz-Thouless (KT) type phase driven by two-valued vortices with the 
transition around $T_{\mathrm{f}}$ \cite{KM}. Slow dynamics associated with bound vortex like defects may 
cause non-ergodicity below a KT type transition into a critical state with a finite $\xi$. 

Other possibilities include the critical behavior toward an AF transition below $T=0.35$ K \cite{Fujimoto}. 
However, it is not yet clear whether such a low-$T$ AF order with geometrical frustration could be robust 
against nonmagnetic impurities of the concentration up to 30 \%. 

Finally, a quantum spin liquid phase is also an interesting candidate. For triangular AFMs, 
Imada {\it et al}. discuss a gapless spin liquid phase through the theoretical calculations based on the 
Hubbard model \cite{SL1}. However, it is not yet clear whether the formation of such a liquid state may 
lead to the $T^2$ dependence of the specific heat. 

To conclude, the Zn substitution for Ni in NiGa$_2$S$_4$ leads to the abrupt loss of the coherence by the 
substitution less than 1 \%. However, the loss is not complete and the robust low temperature
behavior, i.e. the constant susceptibility, the $T^2$ dependent specific heat, and their scaling behavior 
with the Weiss temperature indicate the existence of Nambu-Goldstone mode of the gapless linearly 
dispersive type. The absence of either conventional magnetic order or bulk spin freezing suggests that 
the ground state has a novel symmetry breaking.

We acknowledge Robin T. Macaluso and Julia Y. Chan for sharing their information on the neutron 
diffraction results. We also thank H. Tsunetsugu, H. Kawamura, S. Fujimoto and M. Imada for valuable 
discussions. This work was supported in part by Grants-in-Aid for Scientific Research from JSPS and for 
the 21st Century COE ``Center for Diversity and Universality in Physics'' from MEXT of Japan, and by the 
Tokuyama Foundation.

\end{document}